\documentclass[twocolumn,amsmath,amssymb,prl,superscriptaddress,longbibliography]{revtex4-1}

\usepackage{titlesec}
\usepackage[colorlinks,bookmarks=false,citecolor=blue,linkcolor=red,urlcolor=blue]{hyperref}
\usepackage{epsfig}
\usepackage{color}
\usepackage{graphicx} 
\usepackage{dcolumn}  
 \usepackage{amsmath}
\usepackage{epstopdf}

\newcommand{\be}{\begin{equation}}
\newcommand{\ee}{\end{equation}}

\begin{document}

\title{Strong effect of hydrogen order on magnetic Kitaev interactions in H$_3$LiIr$_2$O$_6$}

 \author{Ravi Yadav}
 \affiliation
 {Institute for Theoretical Solid State Physics, IFW Dresden, Helmholtzstr. 20, 01069 Dresden, Germany}
 
 \author{Rajyavardhan Ray}
 \affiliation
 {Institute for Theoretical Solid State Physics, IFW Dresden, Helmholtzstr. 20, 01069 Dresden, Germany}
\affiliation
{Dresden Center for Computational Materials Science (DCMS), TU Dresden, 01062 Dresden, Germany }
 
 \author{Mohamed S. Eldeeb}
 \affiliation
 {Institute for Theoretical Solid State Physics, IFW Dresden, Helmholtzstr. 20, 01069 Dresden, Germany}
  
 \author{Satoshi Nishimoto}
 \affiliation
 {Institute for Theoretical Solid State Physics, IFW Dresden, Helmholtzstr. 20, 01069 Dresden, Germany}
  \affiliation{Department of Physics, Technical University Dresden, 01062 Dresden, Germany}
 
 \author{Liviu Hozoi}
 \affiliation
 {Institute for Theoretical Solid State Physics, IFW Dresden, Helmholtzstr. 20, 01069 Dresden, Germany}
 
 \author{Jeroen van den Brink}
 \affiliation
 {Institute for Theoretical Solid State Physics, IFW Dresden, Helmholtzstr. 20, 01069 Dresden, Germany}
 \affiliation{Department of Physics, Technical University Dresden, 01062 Dresden, Germany}

\begin{abstract}
Very recently a quantum liquid was reported to form in H$_3$LiIr$_2$O$_6$, an iridate proposed to be a close realization of the Kitaev honeycomb model.
To test this assertion we perform detailed quantum chemistry calculations to determine the magnetic interactions between Ir moments.
We find that 
weakly bond dependent ferromagnetic Kitaev exchange dominates over other couplings, but still is substantially lower than in Na$_2$IrO$_3$. 
This reduction is caused by the peculiar position of the inter-layer species: removing hydrogen cations next to a Ir$_2$O$_2$ plaquette increases the Kitaev exchange by more than a factor of three on the corresponding Ir-Ir link.
Consequently any lack of hydrogen order will have a drastic effect on the magnetic interactions and strongly promote spin disordering. 
\end{abstract}

\maketitle

{\it Introduction.\, }
Quantum spin liquids (QSL's) are states of matter that cannot be described by the broken symmetries
associated with conventional magnetic ground states \cite{balents10}.  Whereas there is a rich
variety of mathematical models that exhibit QSL behavior, finding materials in which a QSL
state is realized is an intensely pursued goal in present day experimental condensed-matter physics
\cite{Shimizu03,yamashita10,han12}.
Of particular interest is the Kitaev Hamiltonian on the honeycomb lattice \cite{Kitaev06}, which
is a mathematically well-understood two-dimensional model exhibiting various topological QSL
states.
Its remarkable properties include protection of quantum information and the emergence of Majorana
fermions \cite{Kitaev06,Pachos12}.

The search to realize the Kitaev model of effective spin-1/2 sites on the honeycomb lattice
was mainly centered until recently on honeycomb iridate materials \cite{jackeli09,chaloupka10}
of the type $A_2$IrO$_3$, where $A$ is either Na or Li. Also of interest is ruthenium trichloride,
RuCl$_3$ \cite{plumb14},
for which Raman and neutron scattering measurements suggest that this $4d^5$ halide honeycomb
system is close to the Kitaev limit \cite{sandilands15,banerjee16,banerjee17}.
However, in all these compounds long-range magnetic order develops at low temperatures, for all known
different crystallographic phases \cite{Singh10,Ye12,choi12,Takayama14,Modic14};
computational investigations suggest that the QSL regime is preempted by sizable Heisenberg residual
couplings, nearest-neighbor (NN) or of longer-range nature \cite{Kimchi11,rau14,Katukuri14,Nishimoto16}.
Very recently, however, it was reported that a Kitaev QSL state of pseudospin-1/2 moments forms in the
honeycomb iridate H$_3$LiIr$_2$O$_6$.
In particular, this material does not display magnetic ordering down to 0.05 K, in spite of magnetic
interaction energies in the range of 100 K~\cite{Kitagawa18}.

Here we present results of quantum chemistry electronic-structure computations for the NN magnetic
interactions between Ir moments and compare these to other honeycomb iridates.
This is done for the crystal structure recently proposed on the basis of x-ray diffraction data~\cite{Kitagawa18}
and also for atomic positions optimized by density-functional calculations.
We find that the Kitaev exchange $K$ is $\sim$10 meV, substantially smaller than in Na$_2$IrO$_3$
\footnote{In $\alpha$-Li$_2$IrO$_3$ peculiar structural distortions cause $J\!>\!K$ for one set of Ir-Ir links 
\cite{Nishimoto16}.},
but in contrast only weakly bond dependent~\cite{Katukuri14} and, most importantly,
the residual NN Heisenberg $J$ is significantly weaker relative to $K$.
The {\it ab initio} calculations show that the smaller absolute $K$ values are 
related to the peculiar position of the inter-layer species, with a single H site neighboring each O
ion~\cite{Kitagawa18}.
Exact-diagonalization (ED) computations using the quantum chemistry NN couplings augmented with
longer-range Heisenberg interactions show the presence of a QSL state, which is however very
susceptible to the formation of long-range order for only weak longer-range exchange, a situation
that is reminiscent of the $A_2$IrO$_3$ (hyper)honeycomb materials.
In this context, the presence of hydrogen vacancies and stacking faults is very interesting:
removing H cations coordinating the bridging ligands on a Ir$_2$O$_2$ plaquette increases the
Kitaev exchange by more than a factor of three for the corresponding Ir-Ir link.
This suggests that the tendency toward the formation of long-range magnetic order in H$_3$LiIr$_2$O$_6$
is very strongly counteracted by H-ion disorder.

\begin{figure}[b]
\includegraphics[width=0.85\columnwidth]{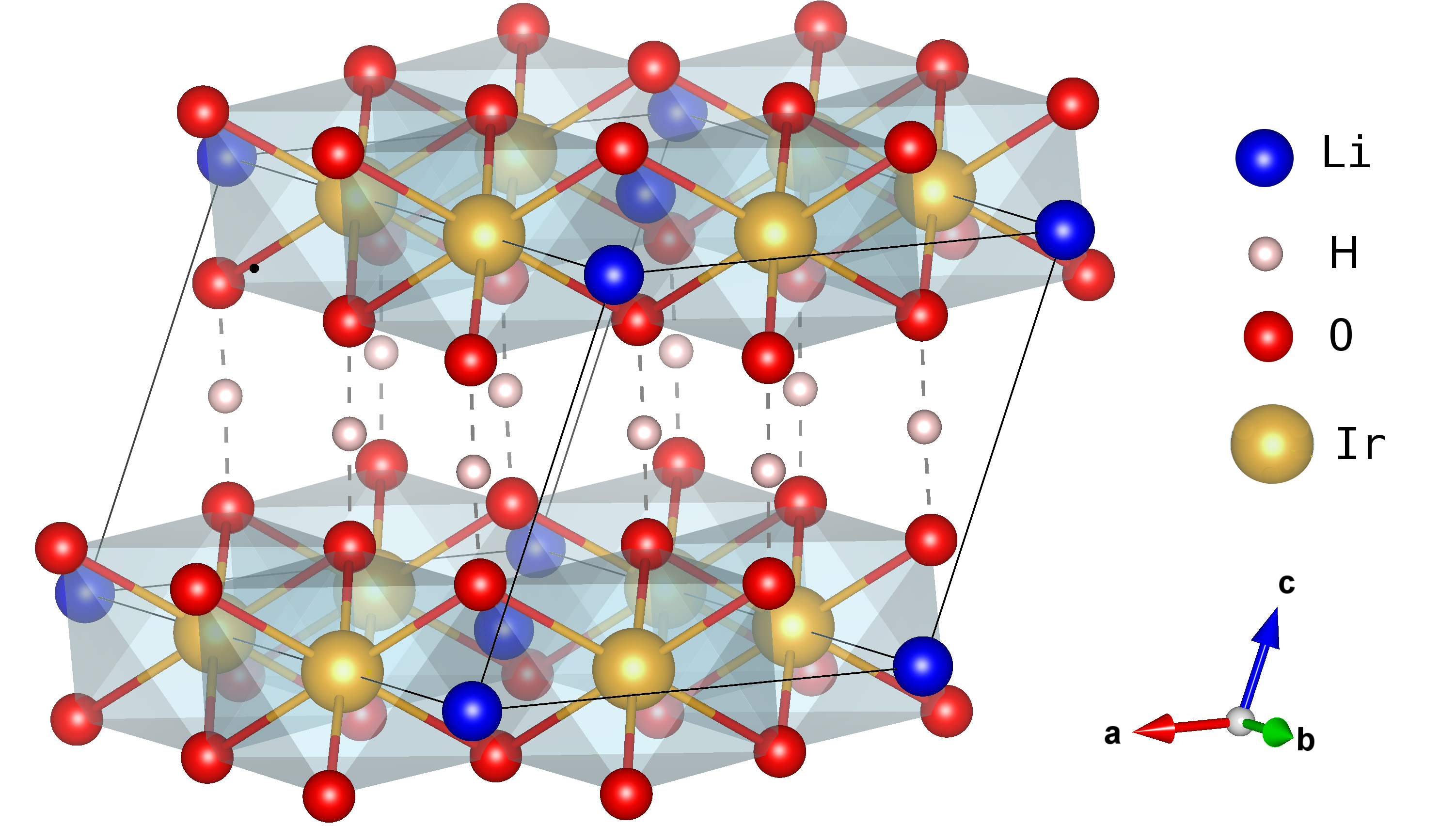}
\includegraphics[width=0.75\columnwidth]{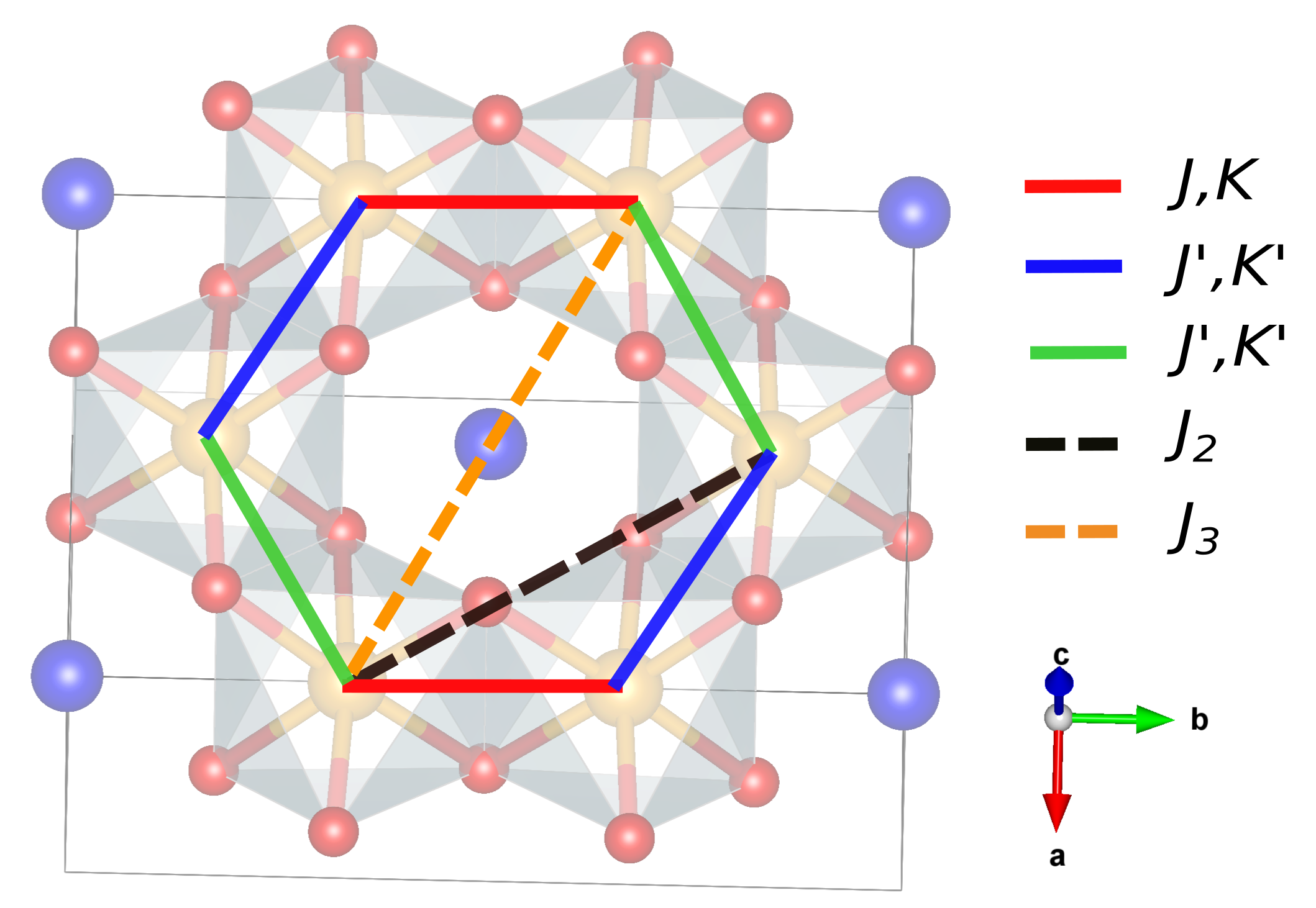}
\caption{
Layered honeycomb network of IrO$_6$ octahedra in H$_3$LiIr$_2$O$_6$.
For the stacking pattern proposed in Ref.\,\cite{Kitagawa18}, interlayer connectivity is realized 
through linear O-H-O links (top); the different types of magnetic couplings on a given hexagonal ring are also shown (bottom).}
\label{fig:structure}
\end{figure}

{\it Basic (electronic) structure.\,}
The spin-orbit-driven Mott insulator H$_3$LiIr$_2$O$_6$ displays a layered structure in which IrO$_6$
octahedra form a planar honeycomb-like network by sharing O-O edges (see Fig.\,\ref{fig:structure}). 
Within this honeycomb lattice, one Li ion is present at the center of each hexagon.
As compared to the related iridate $\alpha$-Li$_2$IrO$_3$, the Li ions sandwiched between two honeycomb
planes are replaced by H species in this recently discovered system.
For the stacking pattern proposed in Ref.\,\cite{Kitagawa18}, inter-layer connectivity is realized
through linear O-H-O links.
The octahedral ligand field splits the Ir $5d$ levels into $e_g$ and $t_{2g}$ states,
with the latter lying at significantly lower energy~\cite{Gretarsson13}.
Given the large $t_{2g}$--$e_g$ splitting,
the leading ground-state configuration is Ir $t_{2g}^5$,
which yields an effective picture of one hole in the $t_{2g}$ sector. 
In the presence of strong spin-orbit coupling, this can be mapped onto a set of fully occupied
$j_{\rm eff}$=3/2 and magnetically active $j_{\rm eff}$=1/2 states~\cite{jackeli09,abragam70,kim08}.
Deviations from a perfect cubic environment may lead to some degree of admixture of these
$j_{\rm eff}\!=\!1/2$ and $j_{\rm eff}\!=\!3/2$ components.

Similar to the parent compound $\alpha$-Li$_2$IrO$_3$, two structurally different types of Ir-Ir links
are present in this system~\cite{Kitagawa18}, which we denote as B1 and B2.
For each of these links, the unit of two NN octahedra displays $C_{2h}$ point-group symmetry, which
then implies a generalized bilinear Hamiltonian of the following form for a pair of pseudospins ${\bf \tilde{S}}_i$
and ${\bf \tilde{S}}_j$\,:
\begin{equation}
 {\cal H}^{(\gamma)}_{ij} =J\, \tilde{\bf{S}}_i \cdot \tilde{\bf{S}}_j
           +K \tilde{S}^\gamma_i \tilde{S}^\gamma_j
           +\sum_{\alpha \neq \beta} \Gamma_{\!\alpha\beta}(\tilde{S}^\alpha_i\tilde{S}^\beta_j +
                                                          \tilde{S}^\beta_i \tilde{S}^\alpha_j)\,,
\label{Eq:ham1}
\end{equation} 
where the $\Gamma_{\alpha\beta}$  coefficients refer to the off-diagonal components of the 3$\times$3 symmetric-anisotropy
exchange matrix, with $\alpha,\beta$\,$\in$\,$\{x,y,z\}$\,\cite{Katukuri14}.
An antisymmetric Dzyaloshinskii-Moriya coupling is not allowed, given the inversion center for
each block of two NN octahedra.
A local Kitaev reference frame is used here, such that for each Ir-Ir link the $z$ axis is perpendicular
to the Ir$_2$O$_2$ plaquette.

{\it Calculation of magnetic interactions.}\,  
To derive magnetic exchange couplings between two NN Ir sites, quantum chemistry calculations were
performed on
embedded clusters having two edge-sharing octahedra
as central region.
To describe the finite charge distribution in the immediate neighborhood, the four adjacent IrO$_6$
octahedra were also included in the calculations,
while the farther solid-state surroundings were modelled by large arrays of point charges fitted to reproduce
the ionic Madelung potential in the cluster region.
In the first step, complete-active-space self-consistent-field (CASSCF) computations\,\cite{Helgaker2000} using an active space of
six $t_{2g}$ orbitals (belonging to two NN Ir sites) and ten electrons
were carried out for an average of the lowest nine singlet and nine triplet states, essentially of
$t_{2g}^5$--$t_{2g}^5$ character.
By construction, $t_{2g}^4$--$t_{2g}^6$ configuration state functions describing intersite $5d$--$5d$
hopping contribute however with finite weight to the CASSCF wavefunctions.
Single and double excitations from the Ir 5$d$ ($t_{2g}$) and bridging-ligand 2$p$ valence shells were
accounted for in the subsequent multireference configuration-interaction (MRCI) calculations\,\cite{Werner88,Knowles92}.
To obtain localized orbitals and select for the MRCI treatement only 5$d$ and 2$p$ orbitals on the
Ir$_2$O$_2$ plaquette, we used a Pipek-Mezey algorithm\,\cite{Pipek89}. A similar computational strategy has been successfully adopted in earlier quantum chemistry studies
\cite{Katukuri14,Bogdanov15,Katukuri14b,Yadav16,bogdanov_12}.

\begin{table}[t]
\caption{Nearest-neighbor magnetic couplings (meV) for the two different bonds B1 and B2 in H$_3$LiIr$_2$O$_6$; results of spin-orbit MRCI
calculations. The structural data provided in Ref.\,\cite{Kitagawa18} was used for this set of calculations.}
\label{Ir_exp}
\begin{tabular}{cccccc}
 \hline
  \hline
     Bond  &\hspace{0.2cm} $\angle$Ir-O-Ir & \hspace{0.2cm}  $K$  & \hspace{0.15cm} $J$ &\hspace{0.1cm}  $\Gamma_{xy}$   & $\Gamma_{yz}$= $-\Gamma_{zx}$ \\
      \hline
      \vspace{-0.1cm}\\
     B2 (3.10\AA{}) &\hspace{0.2cm} $99.8^\circ$  &\hspace{0.2cm} $\textbf{--12.0}$   & \hspace{0.1cm} $\textbf{1.8}$  &\hspace{0.1cm}  $-0.2$  &  $-3.2$ \\
         \vspace{-0.4cm}\\
               \vspace{-0.1cm}\\
     B1 (3.05\AA{}) &\hspace{0.2cm} $99.0^\circ$  & \hspace{0.2cm} $\textbf{--12.6}$   & \hspace{0.1cm} $\textbf{1.5}$  &\hspace{0.1cm}  $-1.8$  &  $-0.7$\\
         \vspace{-0.5cm}\\
              \vspace{-0.1cm}\\
          \hline
         \hline
 \end{tabular}

\end{table}

\begin{table}[b]
\caption{Nearest-neighbor magnetic couplings (meV) for the two different bonds B1 and B2 in H$_3$LiIr$_2$O$_6$;
results of spin-orbit MRCI calculations. The structural data used for this set of calculations
were obtained by DFT lattice optimization.}
\label{Ir_relaxed}
\begin{tabular}{cccccc}
 \hline
  \hline
     Bond  &\hspace{0.2cm} $\angle$Ir-O-Ir & \hspace{0.2cm}  $K$  & \hspace{0.15cm} $J$ &\hspace{0.1cm}  $\Gamma_{xy}$   & $\Gamma_{yz}$= $-\Gamma_{zx}$ \\
      \hline
      \vspace{-0.1cm}\\
     B2 (3.10\AA{}) &\hspace{0.2cm} $98.6^\circ$  &\hspace{0.2cm} $\textbf{--8.8}$   & \hspace{0.1cm} $\textbf{0.8}$  &\hspace{0.1cm}  $-0.4$  &  $-2.7$ \\
         \vspace{-0.4cm}\\
               \vspace{-0.1cm}\\
     B1 (3.05\AA{}) &\hspace{0.2cm} $96.4^\circ$  & \hspace{0.2cm} $\textbf{--6.6}$   & \hspace{0.1cm} $\textbf{0.9}$  &\hspace{0.1cm}  $-2.2$  &  $-0.9$\\
         \vspace{-0.5cm}\\
              \vspace{-0.1cm}\\
          \hline
         \hline
 \end{tabular}

\end{table}

The spin-orbit computations were performed in terms of the low-lying nine singlet and nine triplet states.
The resulting lowest four {\it ab initio} spin-orbit eigenstates were then mapped onto the eigenvectors of the
effective spin Hamiltonian (\ref{Eq:ham1}).
The other 32 spin-orbit states in this manifold involve $j_{\rm eff}\!\approx\!3/2$ to
$j_{\rm eff}\!\approx\!1/2$ excitations and lie at significantly higher energy
\cite{Gretarsson13,Katukuri14}.
The mapping of the {\it ab initio} data onto the effective spin Hamiltonian is carried out following
the procedure described in Refs.\,\cite{Yadav16,Bogdanov15,Yadav17}.
All quantum chemistry computations were performed using the quantum chemistry package {\sc molpro}\,\cite{Molpro12}.

Effective magnetic couplings obtained by such a procedure are listed for the experimental structural
data provided by Kitagawa {\it et al.}\,\cite{Kitagawa18} in Table\,\ref{Ir_exp}.
On both types of Ir-Ir links the Kitaev $K$ is ferromagnetic, with $K\!\approx\!-12$ meV.
The bond `asymmetry' is only 5\% and residual Heisenberg interactions are weak; the ratio $|K/J|$ is
$|K/J|\!>\!6$, which puts the system relatively close to the `pure' Kitaev limit.
The additional exchange anisotropies $\Gamma$ can even exceed $J$ in
magnitude but being frustrating they do not act towards long-range magnetic
order.

To test the stability of the lattice and the effect that small variations of the atomic positions might have on
the magnetic coupling constants,
we also performed structure-optimization calculations in the frame of density-functional theory (DFT).
We employed the Perdew-Burke-Ernzerhof (PBE) variant \cite{pbe96} of the generalized gradient approximation (GGA) with
scalar relativistic corrections as implemented in the FPLO code\,\cite{Klaus99,fplo_web}; the lattice
parameters were fixed to the values derived from x-ray diffraction measurements~\cite{Kitagawa18}.
The DFT relaxed structure corresponds to forces $<$1 meV/{\AA} for each atom.
The most significant difference
between the two sets of atomic positions concerns the location of the ligands
relative to the Ir sites, which then affects somewhat the Ir-O bond lengths and
Ir-O-Ir angles.
For the computationally optimized geometry we also find slight deviations from linear arrangement for
one of the O-H-O links, consistent with the space group symmetry~\cite{Kitagawa18}.
The Ir-Ir distance changes only marginally, $\le$0.3\% (see Supplemental Material
for details\,\cite{supp_mat}). As shown in Table\,\ref{Ir_relaxed}, the Kitaev interactions are slightly smaller
but still dominant in the relaxed structure as well. Also in this case
$|K/J|\!>\!6$ but the differences between the two types of Ir-Ir links are
somewhat larger. Interestingly, in earlier quantum chemistry
calculations for idealized lattice configurations of Li$_2$IrO$_3$ and
Na$_2$IrO$_3$ where all distortions beyond trigonal compression were
neglected it has been found that $J$ tends to zero for Ir-O-Ir bond
angles of approximately 98$^\circ$ \cite{Nishimoto16}. Even if the
Ir-O-Ir bond angles are reduced from 99--100$^\circ$ \cite{Kitagawa18}
to somewhat lower values in the computationally optimized  lattice, we
still obtain nevertheless finite $J$ values by MRCI.

In addition to intersite couplings, we also calculated the Ir 5$d^5$ $g$ factors, by spin-orbit MRCI
computations having a single IrO$_6$ octahedron as reference unit.
The active space in the reference CASSCF calculation includes in this case all 5$d$ Ir orbitals.
The $g$ factors obtained for the $C2/m$ structure of Ref.\,\cite{Kitagawa18} are at the MRCI level
$g_{\alpha\alpha}\!=\!1.83$, $g_{\beta\beta}\!=\!1.98$, and $g_{\gamma\gamma}\!=\!2.30$, where
$g_{\gamma\gamma}$ corresponds to an axis perpendicular to the Ir honeycomb plane and $g_{\beta\beta}$
corresponds to the crystallographic $b$ axis, parallel to the shorter Ir-Ir bonds.
The anisotropy displayed by these components is moderate, far from the highly anisotropic structure
found in the honeycomb Kitaev-Heisenberg system RuCl$_3$\,\cite{Yadav16,kubota15,majumder15}.
Also different from RuCl$_3$ is the fact that $g_{\gamma\gamma}$\,$>$\,$g_{\alpha\alpha}$,\;$g_{\beta\beta}$
in H$_3$LiIr$_2$O$_6$.
Test calculations in which the positive H NN's above and below the reference IrO$_6$ octahedron (six
adjacent H sites in total) are simply removed and their charge redistributed within the embedding 
matrix lead however to values featuring the pattern found in RuCl$_3$, with 
$g_{\gamma\gamma}$\,$<$\,$g_{\alpha\alpha}$,\;$g_{\beta\beta}$,
suggesting that through polarization of the ligand $p$ orbitals the inter-layer cations may significantly
affect the structure of the ${\bar{\bar{\mathbf g}}}$ tensor. Effects of similar nature were found in
square-lattice iridates\,\cite{Bogdanov15}.

\begin{figure}
\includegraphics[width=1.0\columnwidth]{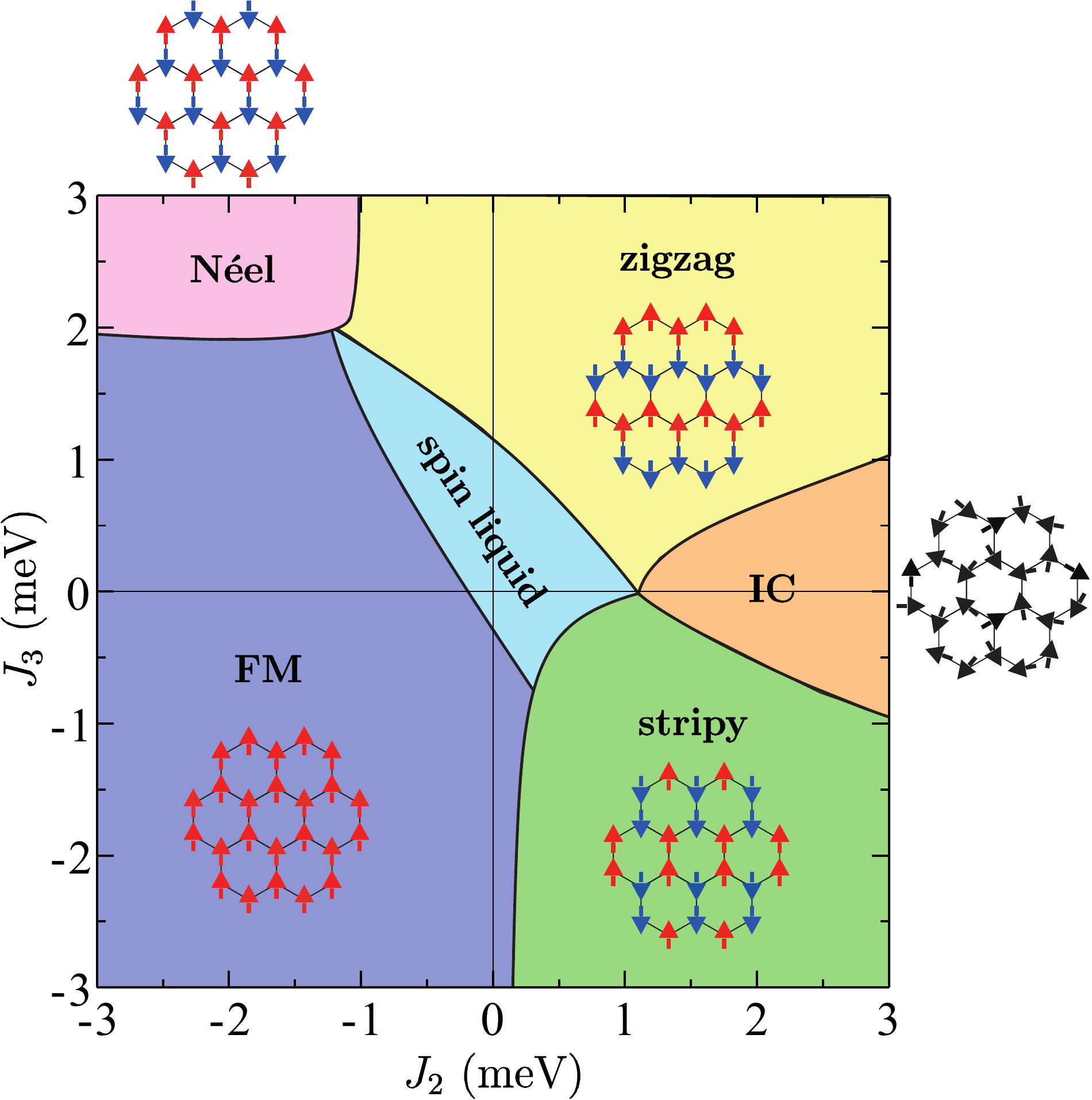}
\caption{
Phase diagram obtained by ED for the effective spin model (\ref{Eq:ham1}).
The {\it ab initio} NN interactions listed in Table\,\ref{Ir_exp} and variable 2nd- and 3rd-neighbor
isotropic couplings $J_2$ and $J_3$ were used.
Schematic spin configurations are also shown.
}
\label{fig:pd_ed_Ir_exp}
\end{figure}

{\it Phase diagram and longer-range interactions.}\,
From the values and bond asymmetry of the NN magnetic interactions, H$_3$LiIr$_2$O$_6$ appears to be
closer to a pure Kitaev model than any other $A_2$IrO$_3$ iridate ($A$=Li,Na) considered so far.
It is known, however, that in the $A_2$IrO$_3$ systems not the residual NN couplings cause at low temperatures the
experimentally observed zigzag ordered state, but the longer-range magnetic
interactions that are present as well, even if the latter can be weak and of the order of 1 meV
\cite{Katukuri14,Majumder18,Yadav16,choi12}.
To test the situation for H$_3$LiIr$_2$O$_6$, we computed a generic phase diagram by using the NN quantum 
chemistry coupling parameters from Table\,\ref{Ir_exp} plus farther-neighbor isotropic Heisenberg $J$'s, second-neighbor ($J_2$)
and third-neighbor ($J_3$).
These calculations were performed as ED for a 24-site cluster with periodic boundary conditions, in
analogy to earlier studies~\cite{chaloupka10,Katukuri14,Yadav16,Nishimoto16}.
The phase boundaries were obtained from the maximum positions in the second derivative of the
ground-state energy.
For a given set of $J_2$ and $J_3$ parameters, the dominant order was determined according to the
wave number ${\bf Q}\!=\!{\bf Q}_{\rm max}$ providing a maximum value of the static structure factor
$S({\bf Q})$.
The QSL state is characterized by a rapid decay of the spin-spin correlations;
a gapless excitation is found here.
More details on this analysis are available in the Supplemental Material\,\cite{supp_mat}.
The resulting phase diagram for variable $J_2$ and $J_3$ is shown in Fig.\,\ref{fig:pd_ed_Ir_exp}.
Four ordered commensurate (FM, N\'eel, stripy, zigzag) phases, an incommensurately ordered (IC), and a QSL phase 
are identified. Representative spin configurations for the ordered phases are also displayed in the figure.

We find that the QSL phase is quickly destabilized by farther-neighbor interactions of Heisenberg type.
If in H$_3$LiIr$_2$O$_6$ the values for $J_2$, $J_3$ are similar to the ones in the $A_2$IrO$_3$ family,
long-range magnetic order of zigzag type is expected for the NN effective couplings computed on the
basis of the crystal structure proposed by Kitagawa {\it et al.}~\cite{Kitagawa18}.
A possibility for QSL ground state remains only when $J_2+J_3\!\lesssim\!1.2$ meV.
If such is indeed realized in H$_3$LiIr$_2$O$_6$, the question arises why the farther-neighbor magnetic
interactions in this material are so much smaller than estimates made for the $A_2$IrO$_3$ systems.

\begin{table}[b]
\caption{
NN magnetic couplings (meV) for bonds B1, B2 in H$_3$LiIr$_2$O$_6$, using structural data from
Ref.\,\cite{Kitagawa18}; results of spin-orbit MRCI calculations where the two H ions next to the
O ligands of a Ir$_2$O$_2$ plaquette were removed and their formal ionic charge redistributed
within the embedding.}
\label{Ir_H_exp}
\begin{tabular}{cccccc}
 \hline \hline
     Bond  &\hspace{0.55cm}  $K$  & \hspace{0.55cm} $J$ &\hspace{0.55cm}  $\Gamma_{xy}$   & \hspace{0.55cm} $\Gamma_{yz}$= $-\Gamma_{zx}$ \\
      \hline
      \vspace{-0.1cm}\\
     B2 (3.10\AA{})  &\hspace{0.55cm} $\textbf{--38.1}$   & \hspace{0.55cm} $5.9$  &\hspace{0.55cm}  $5.0$  &\hspace{0.55cm}  $-11.1$ \\
         \vspace{-0.4cm}\\
               \vspace{-0.1cm}\\
     B1 (3.05\AA{})   & \hspace{0.55cm} $\textbf{--40.0}$   & \hspace{0.55cm} $4.6$  &\hspace{0.55cm}  $7.9$  & \hspace{0.55cm} $-14.0$\\
         \vspace{-0.5cm}\\
              \vspace{-0.1cm}\\        
         \hline
         \hline      
 \end{tabular}
\end{table}

{\it Position of H cations and effect on in-plane interactions.\,}
One peculiar prediction on the quantum chemistry computational side is an enhancement of the Kitaev
interaction for large values of the Ir-O-Ir bond angles \cite{Nishimoto16}.
The latter are 90$^{\circ}$ for cubic edge-sharing octahedra but in most honeycomb compounds become
significantly larger due to trigonal compression of the oxygen cages.
The largest Ir-O-Ir bond angles so far have been actually reported for H$_3$LiIr$_2$O$_6$.
However, given the earlier estimates for $K$ for angles in the range of 98--100$^{\circ}$
\cite{Katukuri14,Nishimoto16}, the $K$ values listed in Tables I and II look surprisingly small.

Comparing the presently known honeycomb iridate compounds, one notable structural difference concerns
the precise position of the inter-layer ionic species:
in Na$_2$IrO$_3$ and $\alpha$-Li$_2$IrO$_3$, for example, the stacking of honeycomb layers is such that
each inter-layer Na or Li site has six O nearest neighbors; on the other hand, in the H-containing
material the available crystallographic data suggest linear inter-layer O-H-O paths with only two O
NN's for each H \cite{Kitagawa18}.
In a simple ionic picture of H$_3$LiIr$_2$O$_6$, the positive H ion next to a given O ligand generates an
axial Coulomb potential that may in principle affect the shape of the O 2$p$ orbitals, thus influencing
the in-plane Ir-Ir superexchange. Obviously, the latter involves the O 2$p$ states.
To test this scenario, we carried out a numerical experiment in which two H cations, in particular, those H nuclei directly coordinating
the bridging O ligands on a given Ir$_2$O$_2$ plaquette, were removed from the atomic fragment treated by quantum chemistry methods but their
associated ionic charge was redistributed within the embedding background. Remarkably, we find in this case an enhancement by a factor
of $\sim$3 of the Kitaev interactions, up to huge values of 40 meV, see Table~\ref{Ir_H_exp}. The other effective magnetic couplings between
NN Ir sites are also enhanced. The larger $J$ values, in particular, indicate that both direct exchange (5$d$--5$d$) and O-mediated superexchange processes contribute to the isotropic coupling constant.
To additionally check how appropriate an ionic representation of the symmetric linear O-H-O links is,
we also derived effective coupling parameters for embedded clusters in which the H NN's of the bridging
O sites (one H cation next to each of the bridging ligands) were represented as 1+ point charges.
The results obtained for this material model show only minor differences as compared to the case in
which basis functions are used for the H species (see Supplemental Material\,\cite{supp_mat}), indicating that an ionic
picture constitutes a rather good approximation for linear O-H-O links and median H positions.
In other words, the strong reduction of the in-plane effective couplings for stacking implying linear
O-H-O groups is mainly related with the destructive effect of the H-cation Coulomb potential on the
Ir-O-Ir superexchange.
Such anisotropic, axial fields are not present when the ligands have several inter-layer adjacent sites
as in $\alpha$-Li$_2$IrO$_3$ and Na$_2$IrO$_3$.

{\it Conclusions.\,}
From the calculations and discussion above it is clear that the configuration of hydrogen cations next
to a Ir-Ir link very strongly affects the magnetic interaction on that link.
Consequently structural hydrogen disorder will introduce very strong magnetic bond disorder.
Any form of hydrogen disorder thereby counteracts the tendency to form long-range ordered states that
are driven by longer-range magnetic couplings.
Experimental investigations into the role of hydrogen disorder on the formation of a spin-liquid
state in this material would therefore be of prime interest to disentangle the effects of NN
Kitaev interactions, that drive the formation of a topological spin-liquid state, and the effect of
hydrogen disorder which induces strong local spin disorder.
In this respect an experimental study of the magnetic and structural properties of H$_3$LiIr$_2$O$_6$
as a function of hydrogen concentration might provide valuable insights.

 {\it Note:\,} 
After finalizing this manuscript we became aware of other computational study on
H$_3$LiIr$_2$O$_6$ using a combination of DFT and non-perturbative ED\,\cite{Li18}.
The effective magnetic couplings and the trends induced by disorder that are reported in that investigation are consistent with our findings.
 


{\it Acknowledgement.\,}
We acknowledge Vamshi M. Katukuri, Klaus Koepernik, and Manuel Richter for helpful discussions and Ulrike Nitzsche for technical support.
This work was supported by the DFG through SFB 1143.
RY and LH also thank the High Performance Computing Center (ZIH) of TU Dresden for access to computational facilities.
RR acknowledges financial support from the European Union (ERDF) and the Free State of Saxony via
the ESF project 100231947 (Young Investigators Group ``Computer Simulations for Materials Design'' -- CoSiMa).

\end{document}